\documentclass[twocolumn]{aa}
\usepackage{graphicx}
\usepackage{epsfig}
\begin{document}

\title{Emission line profiles as a probe of physical conditions in planetary nebulae}

\author{Yong Zhang}

\offprints{Y. Zhang}

\institute{ Department of Physics, University of Hong Kong, Hong Kong \\
\email{zhangy96@hku.hk} }

   \date{Received; accepted }

\abstract
{}
{ We present an analysis of physical conditions in planetary
nebulae (PNe) in terms of collisionally-excited line (CEL) and 
optical-recombination line (ORL) profiles. We
aim to investigate whether line profiles could be used to study
 the long-standing CEL/ORL abundance-discrepancy problem in
nebular astrophysics.  }
{Using 1D photoionization
models and their assumed velocity fields, we simulate the line
profiles of various ionic species.
We attempt to use our model to account for the
observed  CEL and ORL profiles. As a case study we present a 
detailed study of line profiles of the low-excitation 
planetary nebula (PN) IC~418.}
{Our results show that the profiles
of classical temperature and density diagnostic lines,
such as [O~{\sc iii}] $\lambda\lambda$4363,5007, [S~{\sc ii}]
$\lambda\lambda$6716,6731, and [Ar~{\sc iv}] $\lambda\lambda$4711,4740,
provide a powerful tool to study nebular temperature and
density variations. The method enables the CEL/ORL 
abundance-discrepancy problem to be studied more rigorously
than before. A pure photoionization
model of a chemically-homogeneous nebula seems to explain 
the observed disagreements in the profiles for the [O~{\sc iii}] 
$\lambda4363$ and the
$\lambda5007$ lines, but cannot account for the differences
between the [O~{\sc iii}] CELs and the O~{\sc ii} ORLs.
We also investigate the temperature and density variations in the velocity
space of a sample of PNe, which are found to be insignificant.  }
{}

\keywords{planetary nebulae: general --- line: profiles }

\authorrunning{Y. Zhang}

\titlerunning{Line profiles of PNe}

\maketitle

\section{Introduction}

As descendants of AGB (asymptotic giant branch) stars, planetary nebulae
(PNe) serve as crucial probes of the chemical evolution of galaxies and
stellar nucleosynthetic processes. However, the physical conditions inside
PNe are not yet completely understood. An important problem
in nebular astrophysics is that the heavy-element abundances measured
from collisionally-excited lines (CELs) are systematically lower than those
determined from optical-recombination lines (ORLs)
(see Liu \cite{liu06a} for a recent review).
The most extreme known case is that of Hf~2-2 which has an
abundance discrepancy-factor (ADF) of about 70 
(Liu et al. \cite{liubarlow06}).
Possible explanations include electron temperature, electron density, and
chemical-abundance inhomogeneities (Peimbert \cite{peimbert67};
Viegas \& Clegg \cite{viegas94}; Liu et al. \cite{liubarlow00}),
implying that PNe have complex physical conditions.

Many efforts have been made to develop new methods to probe
nebular physical conditions. The traditional methods of using strong
forbidden lines as plasma diagnostics suffer some disadvantages.
Because of their low critical densities ($\la10^5$~cm$^{-3}$), 
CELs are strongly suppressed by collisional de-excitation in high-density
regions, and cannot be used to probe the physical conditions
inside dense clumps within the nebulae.
Given that the fluxes of CELs have an exponential dependence
on electron temperature, they also cannot be used to probe extremely-cold
nebular regions where strong ORLs may arise. Improvements in the quality
of observations and increasingly-accurate atomic data are allowing
 us to use ORLs as plasma diagnostics of PNe.
Zhang et al. (\cite{zhang04}) presented a method to obtain nebular electron temperature and
density simultaneously using hydrogen-recombination spectra. Applying
the method to 48 Galactic PNe, they found that the densities derived
from hydrogen-recombination spectra were generally higher than those derived
 from forbidden line ratios and that the temperatures deduced from
hydrogen-recombination spectra, $T_{\rm e}$(H~{\sc i}), were systematically lower than those
derived from the [O~{\sc iii}] nebular-auroral line-ratios,
$T_{\rm e}$([O~{\sc iii}]), suggesting
that temperature and density variations are generally present in PNe.
Zhang et al. (\cite{zhang05a,zhang05b}) used He~{\sc i} recombination 
line-ratios to
determine nebular electron temperatures, $T_{\rm e}$(He~{\sc i}), and found that
$T_{\rm e}$(He~{\sc i}) $<T_{\rm e}$(H~{\sc i}), supporting the
hypothesis that some cold hydrogen-deficient clumps are embedded in diffuse
gas with a ``normal'' chemical composition, as proposed by 
Liu et al. (\cite{liubarlow00})
to explain the CEL/ORL abundance discrepancies. Some researchers 
presented nebular temperatures and densities
derived from faint O~{\sc ii} recombination-lines 
(Liu \cite{liu06b}; Peimbert \& Peimbert \cite{peimbert05}, and references therein).
Their results suggest that
O~{\sc ii} recombination-lines may arise from colder and denser regions
compared to the diffuse nebular gas.

Line profiles of PNe are of primary importance to investigate nebular
physical conditions. 
To understand nebular dynamical evolution and geometry,
Gesicki et al. (\cite{gesicki03}) and
Morisset \& Stasi\'{n}ska (\cite{morisset06a}) studied velocity profiles of a 
few strong emission lines, in terms of 1D and 3D photoionization models.
A catalog of emission line profiles for PNe with various morphologies
was presented by Morisset \& Stasi\'{n}ska (\cite{morisset06b,morisset08}).
Emission line profiles can also provide an opportunity to test whether CELs
and ORLs originate in different zones.
To complete such a test, 
nebular spectroscopic data of high resolution and 
signal-to-noise are required. Ruiz et al. (\cite{ruiz03}) and 
Peimbert et al. (\cite{peimbert04}) analysed echelle
spectra of the PNe NGC~5307 and NGC~5315. They found that O~{\sc ii} ORLs
 and [O~{\sc iii}] CELs had  similar
widths, implying that both lines originated in the same volume and 
disfavouring the scenario of chemical inhomogeneities. However,
high spectral-resolution observations of two PNe NGC~6153 and NGC~7009 which
have large CEL/ORL abundance discrepancies demonstrated
 that O~{\sc ii} ORLs
are significantly narrower than [O~{\sc iii}] CELs 
(Barlow et al. \cite{barlow06}; Liu \cite{liu06a}),
implying that the ORLs and CELs originate in different zones.
The spectra of NGC~6153 obtained by Barlow et al. (\cite{barlow06}) show that 
the
[O~{\sc iii}] $\lambda4363$ and $\lambda5007$ lines have remarkably-different
profiles, indicating the presence of
large temperature variations within the nebula.

In this paper, we simulate nebular emission line profiles using photoionization
models. Our main objective is to investigate whether line
profiles could be used to study nebular physical conditions and
the problem of ORL/CEL abundance-discrepancies in PNe.
This paper is organized as follows. Section~2 presents our
method. In Sect.~3, we use line profiles to study temperature
and density variation within nebulae and compare with observations.
Discussions and conclusions are presented in Sect.~4.

\section{Method}

\begin{figure}
\begin{center}
\epsfig{file=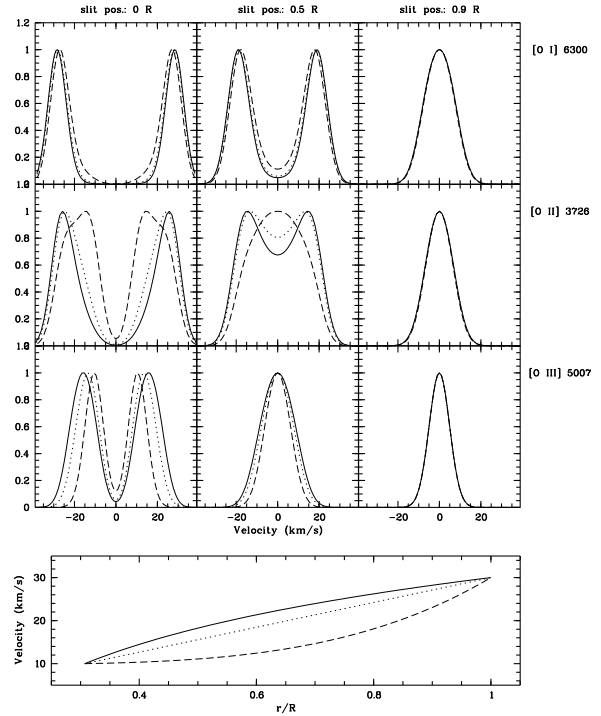,
height=9.5cm, bbllx=34, bblly=101, bburx=545, bbury=727, clip=, angle=0}
\end{center}
\caption{Modelled [O~{\sc i}] (first row), [O~{\sc ii}] (second row),
and [O~{\sc iii}] (third row) line profiles for a nebula observed by
 a long-slit located across the nebular centre (first column), half of the nebular shell
(second column), and the nebular edge (third column).
The peak flux of each line is normalized to unity.
Solid, dashed, and dotted lines correspond to
three different velocity fields
shown in the lower panel.
}
\label{model}
\end{figure}

The basic method is similar to that used by Gesicki et al.
(\cite{gesicki03}) and Morisset \& Stasi\'{n}ska (\cite{morisset06a}).
  We first construct photoionization models using the
CLOUDY code (Ferland et al. \cite{ferland98}). A black-body spectral energy distribution is
adopted for the ionizing stars. The nebulae are assumed to be a
spherical shell with a given radial density distribution.  
CLOUDY calculates the emissivity distribution of the CELs and
ORLs for different ionic species and the thermal structure of nebulae.
Using known slit aperture
and positions and based on an assumed radial-velocity field,
we deduce the integrating line profiles by summing along the line
of sight. 
For our modellings, we assume that the slit width is
far narrower than the nebular radius since high-resolution 
spectra are, in all cases, essential for the study.
The predicted line profiles are then derived by
convolving the broadening caused by temperature,
seeing conditions, and instrument.

Turbulence is not taken into account in
our modelling. Gesicki \& Zijlstra (\cite{gesicki03b}) studied
the expansion-velocity fields of a number of PNe and found that
turbulence might be significant in Wolf-Rayet PNe.
Based on a 3D modelling, 
Morisset \& Stasi\'{n}ska (\cite{morisset06a}), however, argued that
no turbulence is required to explain the observed line profiles
if one considers a nebular-model departure from the assumption of
spherical symmetry. It is therefore safe to assume that the effects of
turbulence, on a normal PN, can be neglected.

Figure~\ref{model} provides an illustration of our modelling of
a medium-excitation PN.  The velocity field is assumed to have
a positive gradient across the nebula. An inspection of Fig.~\ref{model}
shows that the high-ionization lines are narrower than the low-ionization
lines. This is because the high-ionized species are mainly located
within the internal regions where the radial velocity is lower.
The profiles of emission lines arising from species with different ionization
potentials have different dependencies on the velocity fields, and 
provide constraints of the nebular dynamics.
Gesicki et al. (\cite{gesicki96,gesicki98}) and 
Gesicki \& Zijlstra (\cite{gesicki00,gesicki03}) investigated velocity fields
for a sample of PNe by fitting the strong H~{\sc i}, [N~{\sc ii}], and
[O~{\sc iii}] line profiles. They found that the expansion velocities
generally increase outwards, consistent with the predictions of hydrodynamical
calculations.
Deep spectroscopic observations enable weaker lines including CELs
and ORLs which cover a wider range of ionization potentials to be
detected, and tighter constrains on the determination of velocity fields
to be made.
In the deep echelle spectra of the young PN
IC~418, Sharpee et al. (\cite{sharpee04}) detected profiles of a number of 
CELs and ORLs
and found a negative correlation between line widths and ionization
potentials. This is an implication that nebular expansion-velocities
increase outwards.
 Throughout this paper, a monotonously-increasing velocity
field is adopted.

For the model shown in Fig.~\ref{model}, the [O~{\sc ii}] line profiles
are more sensitive to the
velocity field than the [O~{\sc i}] and [O~{\sc iii}] line profiles
because the O$^+$ species are the dominant ionization-state for the
modelled medium-excitation nebula. Figure~\ref{model} also shows that the line
widths decrease when the slit is placed towards the nebular edge.
This tendency
is particularly evident for low-ionization lines originating in
external nebular regions. 
For a given expanding shell, this tendency is present because the velocity 
projected along the line of sight decreases from nebular centre to edge, and 
velocity broadening becomes less important with distance from the nebular 
centre.
At the outer edge of nebulae, thermal broadening is dominant. Therefore,
the emission
lines originating in the outer edge of nebula have a Gaussian  profile
which is essentially independent of expansion velocity and only depends
on electron temperature and atomic mass, as shown in Fig.~\ref{model}.

\section{Analysis}

\subsection{Emission line profiles as a probe of temperature variations}

\begin{figure}
\begin{center}
\epsfig{file=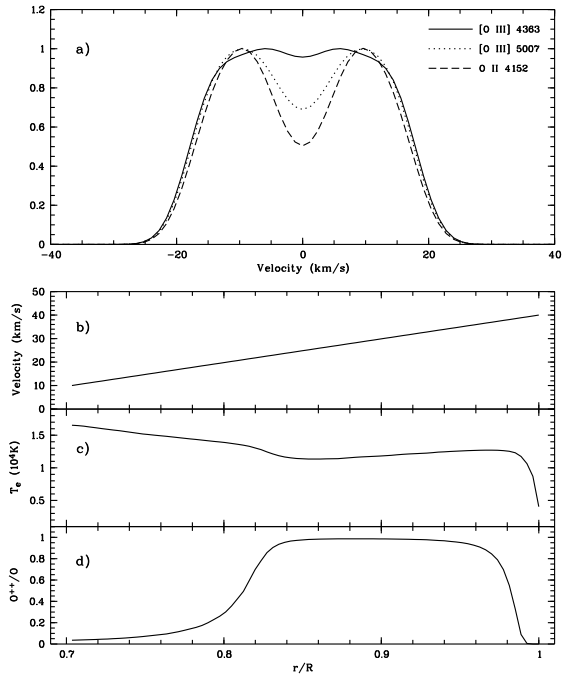,
height=9cm, bbllx=46, bblly=184, bburx=538, bbury=755, clip=, angle=0}
\end{center}
\caption{Modelled results illustrating the effect of temperature variations
on line profiles: a) the predicted line profiles of [O~{\sc iii}] $\lambda4363$,[O~{\sc iii}] $\lambda5007$, and O~{\sc ii} $\lambda4152$; b) the assumed
velocity field; c) the radial temperature distribution; d) the radial
O$^{2+}$/O abundance-ratio distribution.
}
\label{t_flu} \end{figure}

\subsubsection{Modellings}

Two scenarios were proposed to account for the ORL/CEL abundance
discrepancies: 1) temperature fluctuations within the nebula
(Peimbert \cite{peimbert67}); 2) the two-component nebular model with 
extremely-cold
hydrogen-deficient inclusions embedded in the diffuse nebula
(Liu et al. \cite{liubarlow00}).
The investigation of nebular temperature structure is vital for the solution
of this problem. 
To quantitatively characterize the
temperature variations within PNe, Peimbert (\cite{peimbert67}) introduced the
average temperature $T_0$ and the mean square temperature
fluctuation parameter $t^2$, defined by
\begin{equation}
T_0=\frac{\int T_{\rm e}N_{\rm e}N_idV}{\int N_{\rm e}N_idV},
\end{equation}
and
\begin{equation}
t^2=\frac{\int (T_{\rm e}-T_0)^2N_{\rm e}N_idV}{T_0^2\int N_{\rm e}N_idV},
\end{equation}
respectively, where  $N_i$ is the number density of the ionic species
used to determine the electron temperature. For a nebula with small
temperature fluctuations, $T_0$ and $t^2$ can be deduced by comparing
$T_{\rm e}$([O~{\sc iii}]) and $T_{\rm e}$(H~{\sc i}). 
By studying a large number of PNe, Zhang et al. (\cite{zhang04})
 found, however, that the average t$^2$-value of the nebulae 
was higher than predicted by photoionization models. This is probably
evidence of the presence of extremely-cold ($\la1000$\,K) clumps
in which ORLs are significantly enhanced and CELs are barely excited.
To study nebular temperature fluctuations,
Rubin et al. (\cite{rubin02}) obtained a map of the
$T_{\rm e}$([O~{\sc iii}]) of the PN NGC~7009
on the basis of high spatial-resolution Hubble Space Telescope images.
The spatial variation of $T_{\rm e}$([O~{\sc iii}]) in a number
of PNe was also studied by Krabbe et al. (\cite{krabbe05}). 
They found that the value of $T_{\rm e}$([O~{\sc iii}]) remains almost
uniform across the PNe. Nevertheless, their results cannot completely rule
out the scenario of temperature fluctuations
as the cause of ORL/CEL discrepancies
since their measurements of $T_{\rm e}$([O~{\sc iii}]) are an average
along each line of sight and provide only a lower limit to
$t^2$. A comparison of line profiles may provide an unique way to estimate the
temperature fluctuations along the line of sight.

We study the effect of temperature variations across the nebula on
the profiles of the [O~{\sc iii}] $\lambda4363$ and $\lambda5007$ lines.
Figure~\ref{t_flu} shows the modelled results of a high-excitation PN in which
O$^{2+}$ is the dominant ionization state. This model assumes that
the hydrogen density in the nebula
is $n({\rm H})=10^4$\,cm$^{-3}$ and that
the central star has an effective temperature of 150,000\,K and
a luminosity of $4\times10^{37}$\,erg\,s$^{-1}$. As shown in Fig.~\ref{t_flu},
we assume that the
velocity field follows a Hubble law.
The slit is assumed to be midway between the central star and outer edge of
the nebula. For the models used throughout this paper, the average
chemical composition of the Galactic PNe (Kingsburgh \& Barlow \cite{kingsburgh94})
is assumed. 
As shown in Fig.~\ref{t_flu}, the predicted electron temperature
first decreases along the radius and then slightly increases at the outer
edge of the nebula. The low-temperature regions are tightly associated with
the O$^{2+}$/O abundance ratio. This is a natural consequence of
the relatively strong [O~{\sc iii}] nebular lines that
provide the dominant means of cooling in this PN.

Figure~\ref{t_flu} shows that the variations of [O~{\sc iii}]
$\lambda4363$ and $\lambda5007$ line profiles are very sensitive to
nebular thermal structure, and thus provide an opportunity
to probe
nebular temperature variations. In the low-velocity regions where the electron
temperature is higher than the average value, the
[O~{\sc iii}] $\lambda4363$ line has a higher normalized intensity
 than the  [O~{\sc iii}] $\lambda5007$ line. This is because the auroral 
forbidden line has a higher excitation energy, providing higher weighting to 
high temperature regions.

Based on high spectral resolution ($R=150,000$) observations of the PN NGC~6153,
Barlow et al. (\cite{barlow06}) found that the [O~{\sc iii}]
$\lambda4363$ and $\lambda5007$ lines have different profiles, that is,
the gap between the two peaks of the $\lambda5007$ line is filled in the 
case of the $\lambda4363$ line. This can be explained in terms of our
modelling, as shown in Fig.~\ref{t_flu}.
It is interesting that the $t^2$ value ($\sim0.01$) required
to match the observed [O~{\sc iii}] $\lambda\lambda4363,5007$ line profiles
is much lower than that derived from the H~{\sc i} Balmer Jump and
[O~{\sc iii}] lines ($t^2=0.045$; Liu et al. \cite{liubarlow00}). This appears
to support the idea that the two-abundance nebular-component model 
accurately describes the observations, to discount the theory that
temperature fluctuations are the cause of the CEL/ORL abundance dichotomy.
These conclusions are, however, arguable since the analysis of [O~{\sc iii}]
$\lambda\lambda4363,5007$ line profiles can provide
only a lower limit to the temperature variations along the line of sight
(see below).

\subsubsection{Temperature variations in the velocity space} \label{te}

\begin{figure}
\begin{center}
\epsfig{file=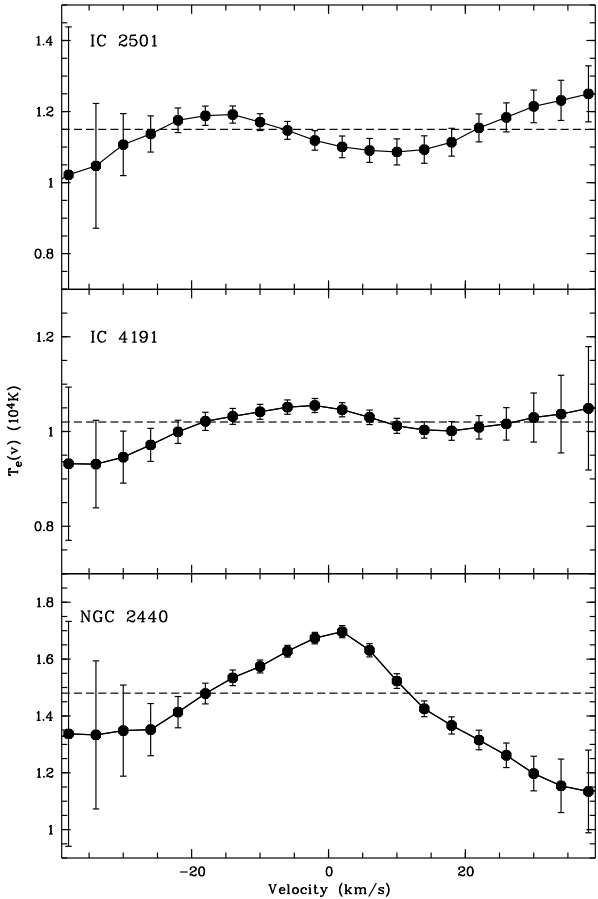,
height=12cm, bbllx=38, bblly=52, bburx=513, bbury=771, clip=, angle=0}
\end{center}
\caption{Electron-temperature distribution in the velocity space
of PNe, IC~2501, IC~4191,
and NGC~2440. The dashed lines represent the values of $T_{0,v}$.
}
\label{t_v}
\end{figure}

To study the temperature variations, we define the
average temperature $T_{0,v}$ and the mean square temperature fluctuation
parameter  $t^2_v$ in velocity space by
\begin{equation}
T_{0,v}=\frac{\int T_{\rm e}(v)I_v({\rm H}\beta)dv}
{\int I_v({\rm H}\beta)dv},
\end{equation}
and
\begin{equation}
t^2_v=\frac{\int [T_{\rm e}(v)-T_{0,v}]^2I_v({\rm H}\beta)dv}{T_{0,v}^2\int I_v({\rm H}\beta)dv},
\end{equation}
where $v$ is the velocity along the line of sight, and
$T_{\rm e}(v)$ and $I_v({\rm H}\beta)$ are
 the electron temperature deduced from the
forbidden line intensity ratio and the intensity of H$\beta$
($\propto N_{\rm p}N_{\rm e}$) at a given velocity, respectively.
The integrations in the two equations are over velocity space.
As defined by Eqs.~(3) and (4), $T_{0,v}$ and $t^2_v$
characterize temperature fluctuations along the line of sight
and are similar to those introduced to
explain the $T_{\rm e}$([O~{\sc iii}])/$T_{\rm e}$(H~{\sc i})
discrepancy, which are given in Eqs.~(1) and (2). In Eqs.~(3) and (4), 
we can use $I_v({\rm X}^{i+})$, which is 
the intensity of a CEL used to determine $T_{\rm e}(v)$,
as weights instead of $I_v({\rm H}\beta)$. However, if 
significant density variations are present along the line of sight, 
it may be problematic to use
 $I_v({\rm X}^{i+})$ since its density 
dependences differ in high- and low-density regions.

Using Eqs.~(3) and (4), we calculated $T_{0,v}$ and $t^2_v$ for three
PNe, IC~2501, IC~4191, and
NGC~2440.  Our data were acquired  using the Las Campanas Observatory (LCO) Baade
6.5m telescope with the MIKE echelle spectrograph, yielding a resolution of
$\sim25,000$
(see Sharpee et al. \cite{sharpee07} for a detailed description of the 
observations).
The results are given in Table~\ref{tab1}. For IC~4191 and NGC~2440
the values of $T_{0,v}$ and $t^2_v$
are measured using the [O~{\sc iii}] $\lambda\lambda4363,5007$ lines.
For the low-excitation PN IC~2501, the slit was placed on the edge of the
compact O$^{2+}$ regions and the thermal broadening provides a
dominant contribution to the profiles of the [O~{\sc iii}] lines. 
For IC~2501 we obtained $T_{0,v}$ and $t^2_v$ using instead the
single-ionization lines [N~{\sc ii}] $\lambda\lambda5754,6678$ lines.
 For the calculations, the reddening
coefficient and the electron densities presented by 
Sharpee et al. (\cite{sharpee07}) were adopted.
As depicted in Fig.~\ref{t_v}, the temperature variations in the velocity
space are minor for the three PNe.

\begin{table}
\caption{ The values$^a$ of $T_{0,v}$ and $t^2_v$.
 \label{tab1}}
\begin{center}
  \begin{tabular}{@{}cccc@{}}
\hline
Object & {IC~2501} & {IC~4191} & {NGC~2440} \\
\hline
$T_0$ (K)     & 7700 &  8600 & 12600 \\
$t^2$         & 0.053&  0.040& 0.077 \\
$T_{0,v}$ (K) & 11500&  10200& 14800 \\
$t^2_v$       & 0.002&  0.001& 0.011 \\
\hline
\end{tabular} 
\begin{description}
\item $^{a}$ For the low-excitation PN, IC~2501, $T_{0,v}$ and $t^2_v$
are deduced using the [N~{\sc ii}] $\lambda\lambda5754,6678$ lines; for
the high-excitation PNe, IC~4191 and NGC~2440, $T_{0,v}$ and $t^2_v$
are deduced using the [O~{\sc iii}] $\lambda\lambda4363,5007$ lines.
\end{description}
\end{center}
\end{table}

Based on the values of
$T_{\rm e}$([O~{\sc iii}]) and $T_{\rm e}$(H~{\sc i}) presented by
Sharpee et al. (\cite{sharpee07}), we determined
$T_0$ and $t^2$ using Peimbert's formulae (Peimbert \cite{peimbert67}).
The calculations were based on the
assumption that the O$^{2+}$ and H$^{+}$ regions are identical,
which is appropriate for our studied PNe since their
O$^{+}$/(O$^{+}$+O$^{2+}$) abundance ratios are consistently lower than 0.1 
(Sharpee et al. \cite{sharpee07}).
The resultant $T_0$ and $t^2$ are given in Table~\ref{tab1}.
If the discrepancy between $T_{\rm e}$([O~{\sc iii}]) and $T_{\rm e}$(H~{\sc i})
is caused by small temperature fluctuations, the $t^2$ measured by
$T_{\rm e}$([O~{\sc iii}]) and $T_{\rm e}$(H~{\sc i}) should be
identical to those measured along the line of sight. However,
Table~\ref{tab1} shows that the values of
$t^2_v$ are consistently lower than those of $t^2$ and the values of
$T_{0,v}$ are consistently higher than those of $T_0$, seemingly
favoring the presence of extremely-cold nebular components where
the [O~{\sc iii}] lines cannot be excited but the H~{\sc i}
recombination spectrum  is significantly strengthened.

This conclusion, however, needs to be treated with caution.
The value of  $t^2_v$ defined by Eq.~(4) represents the
temperature variations in velocity space. We should bear in mind that the
validity of the argument that $t^2_v$ is equal to the mean square
temperature variation parameter along the line of sight
requires two conditions to be satisfied:
a) nebular expansion-velocity dominates the broadening of line profiles;
b) the expansion-velocity is monotonic along the nebular
radius, as found in most of PNe. 
Otherwise, $t^2_v$ provides only a lower limit to the temperature fluctuations
along  the line of sight.  In any case, the comparison
of  $t^2_v$ and $t^2$ provides an opportunity to investigate the presence
of extremely-cold components. This paper is beginning such a project.
Higher-resolution spectroscopic data of PNe with a high expansion velocity
will be invaluable for this study.

\subsubsection{ ORL profiles versus CEL profiles}

Figure~\ref{t_flu} also depicts that if temperature variations exist,
the O~{\sc ii} lines show a profile that differs from that of the
[O~{\sc iii}] lines, which can be attributed to the fact that
CELs and ORLs weight high- and low-temperature regions, respectively.
In the low velocity regions, the electron temperature is relatively
high. As a result, the O~{\sc ii} lines have a deeper ``gap'' between
the two peaks than the [O~{\sc iii}] lines. In the high velocity
regions which correspond to the line wings, the electron temperature
is slightly higher, depressing recombination lines. Consequently,
the O~{\sc ii} lines are narrower than the [O~{\sc iii}] lines.
The O~{\sc ii} line profiles differ most significantly from those of the 
[O~{\sc iii}] $\lambda4363$ line than of the [O~{\sc iii}] $\lambda5007$
 line, since the [O~{\sc iii}] $\lambda5007$ line has a lower excitation 
temperature than the [O~{\sc iii}] auroral line.

High spectral-resolution observations of PN NGC~7009 show
that the [O~{\sc iii}] $\lambda4363$ CEL is broader than the
O~{\sc ii} ORLs by a factor of about 1.5 (Barlow et al.
\cite{barlow06}). For their observations, a $0.9\arcsec\times0.9\arcsec$
image-slicer was placed on a bright-edge region located $5.6\arcsec$
northwest of the central star.
We attempted to model the observed CEL/ORL width discrepancy. However,
we find that chemically-homogeneous nebula models
are unable to reproduce a CEL/ORL width ratio that is greater
 than 1.2 because our modellings are unable to achieve such large
temperature variations. The modelled
CEL/ORL width ratio is essentially independent of the adopted velocity
field and is insensitive to the assumed density distribution since
the [O~{\sc iii}] $\lambda4363$ CEL and the O~{\sc ii} ORLs both  have
high critical densities. To match the observed CEL/ORL widths properly,
the electron temperature is required to increase sharply outwards, 
which is difficult to achieve using pure photoionization
modelling of gaseous nebulae.
The most direct explanation for the observed large CEL/ORL width discrepancy
is that the O$^{2+}$ CELs and ORLs originate
in different nebular components which have separate kinematic fields.
Liu (\cite{liu03}) proposed that evaporating planetesimals within the
nebulae might produce
cold H-deficient inclusions where ORLs are greatly enhanced and CELs are
depressed. If this is the case, these H-deficient components should
move in relatively-stable orbits, and thus have a lower velocity
dispersion than the diffuse nebular gas.
As an alternative, it remains possible that heating by shock waves
in the outer regions may be significant and leads
to a sharp temperature increase in the high-velocity regions.
Furthermore, Stasi\'{n}ska \& Szczerba (\cite{stasinska01})
 showed that
photoelectric heating by dust grains can cause large
temperature fluctuations in the presence of
density inhomogeneities.
However, in the case of a constant dust-to-gas ratio in the entire nebula,
the temperature is more efficiently
enhanced in the inner regions. If the CEL/ORL width discrepancy
is caused by heating by dust, these dust grains must be predominantly located 
in the outer zones of the ionized gas.

\subsection{Emission line profiles as a probe of density variations}

\subsubsection{Modellings}

\begin{figure}
\begin{center}
\epsfig{file=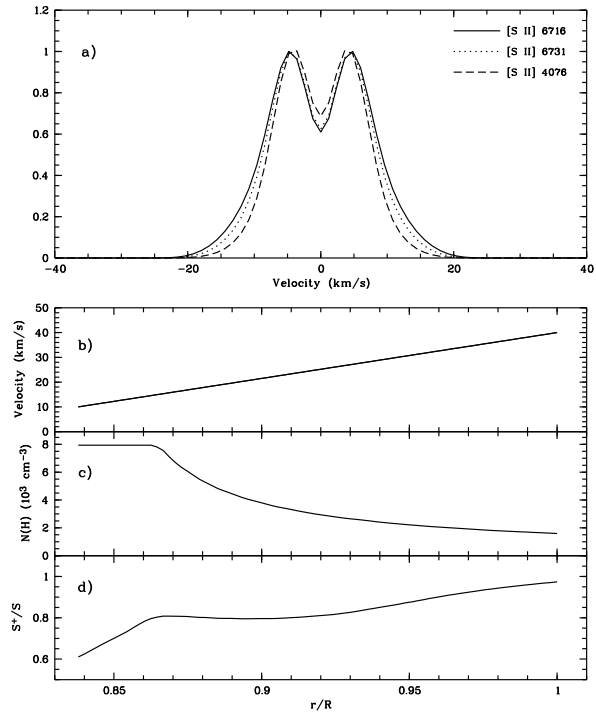,
height=9.5cm, bbllx=46, bblly=184, bburx=538, bbury=755, clip=, angle=0}
\end{center}
\caption{Modelled results illustrating the line profiles for an
outwardly-decreasing density distribution: a) the predicted line profiles
of [S~{\sc ii}] $\lambda6716$, [S~{\sc ii}] $\lambda4731$, and
[S~{\sc ii}] $\lambda4070$; b) the assumed velocity field; c) the radial
density distribution of hydrogen atoms; d) the radial S$^+$/S abundance-ratio
distribution.
}
\label{n_neg} \end{figure}

The determination of nebular density structure is critical
to the investigation of ORL/CEL abundance discrepancies. 
Viegas \& Clegg (\cite{viegas94}) pointed out
that the presence of high density regions could lead to an overestimate of
$T_{\rm e}$([O~{\sc iii}]) and consequently an underestimate
of CEL abundances
since the  $\lambda$5007 nebular line has a far lower critical density than
the [O~{\sc iii}] $\lambda$4363 auroral line and is more significantly
suppressed by collisional de-excitation in-high density regions.
However, Liu et al. (\cite{liubarlow00}) claimed that density inhomogeneities 
can be ruled out 
as the cause of the ORL/CEL abundance discrepancies 
because no correlation between the abundance discrepancies and
critical densities was found. 
Line profiles can  be used to probe density variations along 
the line of sight.

\begin{figure}
\begin{center}
\epsfig{file=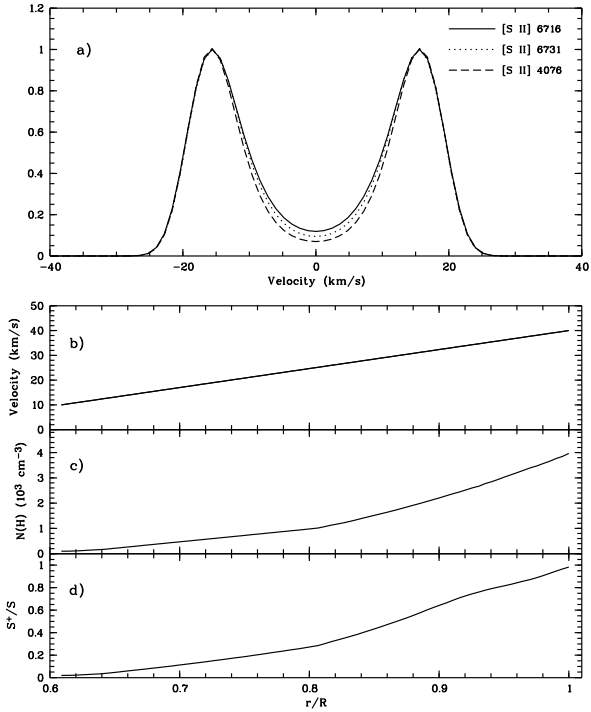,
height=9.5cm, bbllx=46, bblly=184, bburx=538, bbury=755, clip=, angle=0}
\end{center}
\caption{Same as \ref{n_neg} but for an outwardly-increasing density
distribution.
}
\label{n_pos} \end{figure}

The [S~{\sc ii}] $\lambda\lambda6731,6716$ doublet lines are classical
electron-density diagnostics.  We therefore simulate the profiles of the
[S~{\sc ii}] lines under the assumption that density decreases outwards
and increases outwards, as shown in Figs.~\ref{n_neg}
and \ref{n_pos}, respectively. We assume that the ionizing star
has a temperature of 30,000\,K. For the nebulae that we model,
S$^+$ is the dominant ionization-state of sulphur. Figures~\ref{n_neg} and
\ref{n_pos} indicate that the  the profiles of the
[S~{\sc ii}] $\lambda\lambda6731,6716$
doublet lines differ  if density variations are present.
The [S~{\sc ii}] $\lambda6731$ line has a higher critical density of
$\sim4\times10^3$\,cm$^{-3}$ than the $\lambda6716$ line
($\sim1\times10^3$\,cm$^{-3}$), and hence is less-affected by
collisional de-excitation in high-density regions. 
For a density that decreases and increases outwards, the [S~{\sc ii}]
$\lambda6731$ line is then stronger in low- and high-velocity regions, respectively.
As shown in Figs.~\ref{n_neg} and \ref{n_pos},
the profile discrepancy is particularly remarkable when the
[S~{\sc ii}] $\lambda\lambda6731,6716$ lines and the
[S~{\sc ii}] $\lambda4076$ line are compared
because the latter has a much higher
critical density ($\sim10^6$\,cm$^{-3}$). However, we note
 that temperature fluctuations may contribute partially to 
the profile discrepancy between these density diagnostic lines.

Apart from for the density distribution, the basic assumptions are identical
in each model.
Figure~\ref{n_pos} shows a larger separation between the two emission peaks
than in Fig.~\ref{n_neg}. This is because in the case of 
outwardly-increasing density distribution, emission lines weight
the high-velocity regions more highly.

\begin{figure}
\begin{center}
\epsfig{file=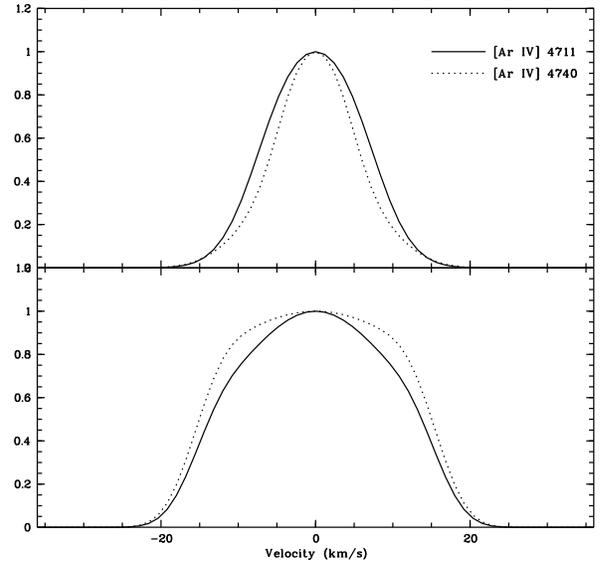,
height=7.5cm, bbllx=62, bblly=324, bburx=538, bbury=755, clip=, angle=0}
\end{center}
\caption{The profiles of the [Ar~{\sc iv}] $\lambda\lambda4711,4740$
lines for outwardly-decreasing (upper panel) and increasing (lower panel)
density distributions.
}
\label{ar} \end{figure}

The analysis methods are applied to other density diagnostics, such
as [O~{\sc ii}] $\lambda\lambda3726,3729$, [Cl~{\sc iii}]
$\lambda\lambda5517,5537$, and [O~{\sc ii}] $\lambda\lambda4711,4740$.
As an example, Fig.~\ref{ar} shows the profiles of the
[Ar~{\sc iv}] $\lambda\lambda4711,4740$ lines for high-excitation
PNe with a stellar temperature of 150\,000\,K under the assumption
of outwardly decreasing and increasing density distributions. 
In the two sorts of density structures,
the relations between the two line profiles are completely the opposite.
The [Ar~{\sc iv}] $\lambda4711$
line has a lower critical density  than
the [Ar~{\sc iv}] $\lambda4740$ line, and
thus weights lower-density regions more highly. 
Hence, if the nebula has a negative
density gradient, the [Ar~{\sc iv}] $\lambda4711$ line weights the
outer regions more, where the expansion velocity is larger, which
produces a broader profile in  the [Ar~{\sc iv}] $\lambda4711$ line compared
to the [Ar~{\sc iv}] $\lambda4740$ line. For a similar reason,
the [Ar~{\sc iv}] $\lambda4711$ line has a narrower profile than
the [Ar~{\sc iv}] $\lambda4740$ line for an outwardly increasing density
distribution. A comparison of emission line profiles for various
ionic species provides tighter constraints on nebular
density structure.

\subsubsection{Density variations in the velocity space}

\begin{figure}
\begin{center}
\epsfig{file=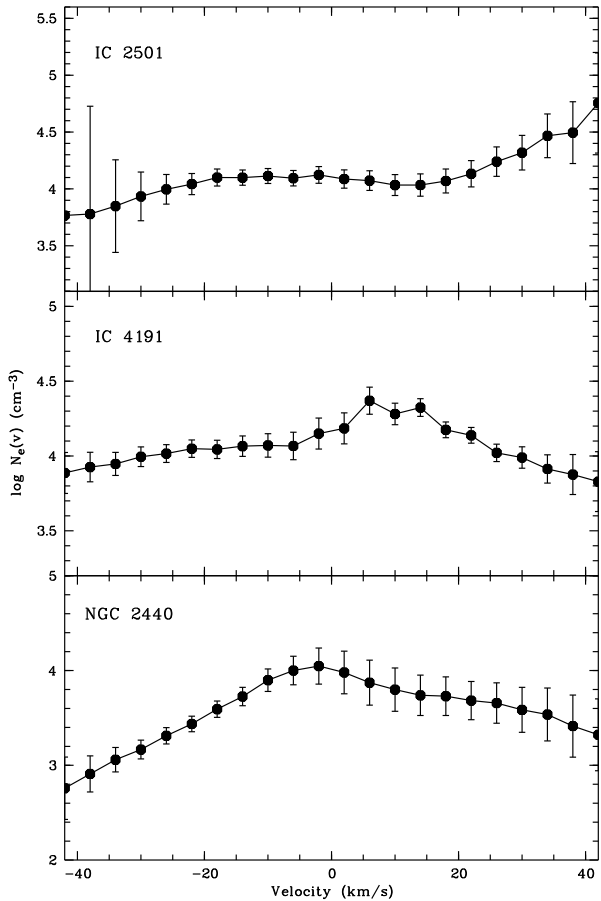,
height=12cm, bbllx=38, bblly=52, bburx=513, bbury=771, clip=, angle=0}
\end{center}
\caption{Electron-density distribution in the velocity space of
the PNe,
IC~2501, IC~4191, and NGC~2440.
}
\label{n_v}
\end{figure}

Using the [S~{\sc ii}] $\lambda\lambda6731,6716$ line profiles,
we calculated the density distribution in velocity space, $N_{\rm e}(v)$,
for the PNe IC~2501, IC~4191, and NGC~2440 
(see Sharpee et al. \cite{sharpee07}
for the observation details). The results are shown in
Fig.~\ref{n_v}. An inspection of the figure clearly indicates that
the density variations in the velocity field are small (less than 1 dex).
However, for the same reasons discussed
in Sect.~\ref{te} for $T_{\rm e}(v)$, density variations in the velocity 
field may have been
smoothed and only represent a lower limit to the nebular density variations
unless line broadening is dominated by the expansion velocity and the velocity
field is monotonic along the nebular radius.

Figure~\ref{n_v} shows that the compact PN IC~2501
has an almost homogeneous density structure  with an average density
of $\sim10^4$\,cm$^{-3}$. The slight increase in density
at 30\,km/s is not real
considering the large errors introduced by the weakness of the
line wings. IC~4191 is a compact and irregular PN. Figure~\ref{n_v} shows that
it has an asymmetrical density structure. A condensation is evident 
around 10\,km/s on the red side,  which has a density about twice
as high as the other nebular regions. We cannot find a
counterpart on the blue side. NGC~2440 clearly shows a negative
density gradient across the nebula. The density at the outer edge of
the nebula is lower by an order of magnitude than at the centre.

We find that the [S~{\sc ii}] $\lambda4076$ line has a 
different profile from the [S~{\sc ii}] $\lambda\lambda6731,6716$
lines for all three PNe.
However, the [S~{\sc ii}] $\lambda4076$ line was not taken into account in
the current analysis because its upper level has a higher excitation
energy than the [S~{\sc ii}] $\lambda\lambda6731,6716$, and hence
we cannot separate the contribution of temperature variations to
 the difference between their profiles.

\subsection{Emission line profiles of IC~418} \label{ic418}

To demonstrate how line profiles could be used to measure nebular
physical conditions, we considered data for IC 418 to complete a 
detailed study of line profiles by attempting to match our models to 
observational data.
The analysis of a larger number of PNe will appear in a separate paper.
High-resolution echelle spectra of the PN IC~418
were obtained in 2001 with
the 4 m Blanco telescope at Cerro Tololo Inter-American Observatory
(Sharpee et al. \cite{sharpee04}), covering a wavelength range from
3500 to 9865\,{\AA}. IC~418 is a low-excitation PN
[$F$([{\rm O}~{\sc iii}]~$\lambda5007)/F({\rm H}\beta)\sim2$] and has an 
approximatively-spherical shell. The observations were carried out using a
$\sim1''$ slit width, yielding a spectral resolution of $\sim30,000$.
The spectrograph slit was aligned north-south.
The central star was placed along a line perpendicular to the spectrograph
slit so that the slit centre was roughly midway
between the central star and outer edges of the nebula.
A large number of ORL and CEL profiles are well resolved.

Hyung et al. (\cite{hyung94}) constructed a photoionization model of IC~418 
that can interpret the UV data fairly well.
For our model, we initially took the basic parameters used by 
Hyung et al. \cite{hyung94}
(distance, properties of the central star, chemical composition,
nebular geometrical structure, etc.),
which were then adjusted slightly to achieve an optimal fit to
optical observations. The density distribution of
IC~418 is a topic of ongoing debate.
Hyung et al. (\cite{hyung94}) presented that the nebula may consist of two 
shells, an inner high-density one and an outer lower-density shell. A 
conflicting
conclusion was obtained by  Gesicki et al. (\cite{gesicki96}) who found that
the density increases smoothly outwards
by fitting the surface-brightness profiles for H$\beta$, [O~{\sc iii}]
$\lambda5007$, and [N~{\sc ii}] $\lambda6584$.
Our model assumed a single shell with a homogeneous density of
$N_{\rm H}=10^4$\,cm$^{-3}$.
Below we show that this issue can be settled by 
 comparing the profiles of density diagnostic lines.

\begin{figure}
\begin{center}
\epsfig{file=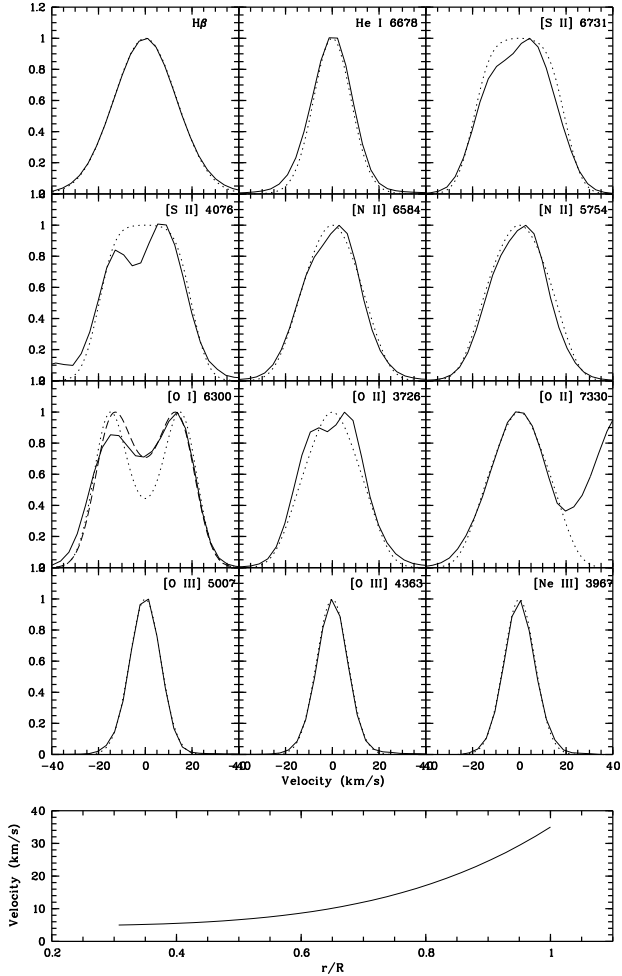,
height=13cm, bbllx=40, bblly=30, bburx=515, bbury=771, clip=, angle=0}
\end{center}
\caption{Observed (solid lines) and predicted (dotted lines) strong emission
lines of IC~418. The assumed velocity field is shown in the lower panel.
For the [O~{\sc i}] $\lambda6300$ line, the dashed line represents
the predicted profile by refined emission distribution, as discussed
in the text.
}
\label{ic418_c}
\end{figure}

Once the photoionization model provides satisfactory fits to the observed
integrated line intensities, we calculate the line profiles by
assuming a velocity field and compare the predicted results
with the observed ones. For the comparison, we used
a number of strong lines from different ionic species, including
H~{\sc i}, He~{\sc i}, [S~{\sc ii}], [N~{\sc ii}], [O~{\sc i}],
[O~{\sc ii}], [O~{\sc iii}], and [Ne~{\sc iii}]. The observed and
predicted line profiles are presented by Fig.~\ref{ic418_c}.
Low-ionization lines have generally broader profiles than
high-ionization lines, which implies that
 the expansion velocity is increasing outwards.
To reproduce the observed line profiles, the acceleration  is required
to increase in the outer regions, as shown in Fig.~\ref{ic418_c}.
For the fitting, no turbulence broadening is needed.
Figure~\ref{ic418_c} shows that our model can account for most of
these strong lines.

\begin{figure}
\begin{center}
\epsfig{file=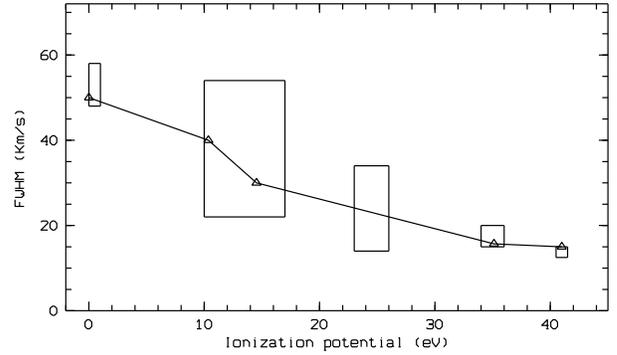,
height=4.8cm, bbllx=38, bblly=106, bburx=516, bbury=389, clip=, angle=0}
\end{center}
\caption{FWHM as a function of the ionic species' ionization potential
for the PN IC~418. The solid line with triangles is the prediction of
our model. The box denotes the zone occupied by the observations of
Sharpee et al. (\cite{sharpee04}).
}
\label{ip}
\end{figure}

Figure~\ref{ip} plots the FWHMs of the lines as a function of ionization 
energy,
as predicted by our model and observed by Sharpee et al.
(\cite{sharpee04}) (see their Fig.~7).
Good agreement between the predictions and the observations
is achieved.
The velocity field produced is in general agreement with the results of
Gesicki et al.
(\cite{gesicki96}), but has a larger outer velocity. Given that
Gesicki et al.
(\cite{gesicki96}) did not consider the [O~{\sc i}] line, their
measured outer velocity is unlikely to be reliable. The sharply-increasing 
velocity in the outer regions is associated with the shock
at the ionization front, which is consistent with the predictions of
hydrodynamic models (e.g. Perinotto et al. \cite{perinotto98}).

The [O~{\sc ii}] $\lambda7330$ line can be explained by our model, while
the observed [O~{\sc ii}] $\lambda3726$ line is broader than the predicted
one. This is unlikely to be due to temperature variations
since the [N~{\sc ii}] $\lambda\lambda5754,6584$ doublet lines is
reproduced well.  The discrepancy, therefore, can be attributed to  density
variations within the PN. The [O~{\sc ii}] $\lambda3726$ line has a
critical density about three times lower than that
of the [O~{\sc ii}] $\lambda7330$ line, and thus originates in low-density
regions. It follows that the outer regions have a lower density than
the inner regions, in contrast to the results of 
Gesicki et al. (\cite{gesicki96}).

For the [O~{\sc i}] $\lambda6300$ line, our model reproduces the
profile in the high-velocity range well.
But in the low-velocity range, the observed flux is much higher than
predicted.  Morisset \& Stasi\'{n}ska (\cite{morisset06a})
also encountered difficulty in fitting the low-velocity range of the 
[O~{\sc i}] $\lambda6300$ line profile.
A possible explanation is that neutral gas is
plentifully abundant within the ionized regions.
This conjecture was supported by Williams et al. (\cite{williams03}), who 
compared absorption- and emission-line spectra of IC~418 and found evidence
 that absorbing gas may be present.
If we enhance the [O~{\sc i}] emissivity in the inner regions by
a factor of $\sim$1.5, an excellent match between the predicted profile and observed one
is achieved, as shown the Fig.~\ref{ic418_c}.

For observations of low-excitation PN,
the slit was placed in the outer edge of the O~{\sc ii} regions where
the projected velocity along the line of sight is quite low.
Therefore, the narrow high-ionization lines, such as [O~{\sc iii}] and
[Ne~{\sc iii}], are dominated by thermal and instrumental broadening.
The [O~{\sc iii}] $\lambda\lambda4363,5007$ lines are therefore
unsuitable for deriving temperature variations in velocity space.

\begin{figure}
\begin{center}
\epsfig{file=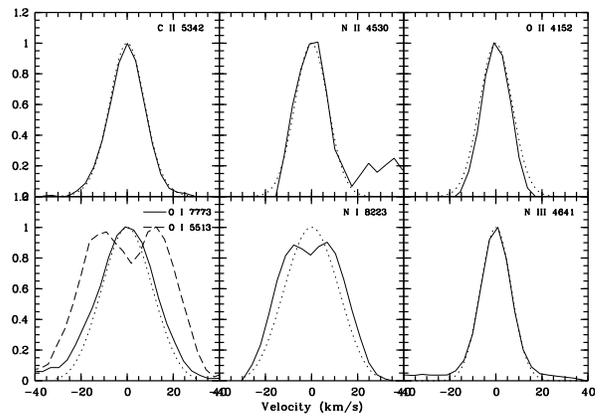,
height=5.5cm, bbllx=59, bblly=453, bburx=515, bbury=771, clip=, angle=0}
\end{center}
\caption{Same as Fig.~\ref{ic418_c} but for recombination lines.
}
\label{ic418_r}
\end{figure}

We compare the predicted ORL profiles with the observations.
Figure~\ref{ic418_r} displays a few examples for such a comparison.
Generally, our model shows  good agreement with observations of the
high-photoionization lines, C~{\sc ii}, N~{\sc ii},
O~{\sc ii}, and N~{\sc iii}. 
IC~418 has a small ORL/CEL abundance discrepancy
[$({\rm O}^{2+}/{\rm H}^+)_{\rm ORL}/({\rm O}^{2+}/{\rm H}^+)_{\rm CEL}=1.3$;
Sharpee et al. \cite{sharpee04}],
and is therefore an unsuitable target to investigate the materials in which
ORLs originate.

Figure~\ref{ic418_r} demonstrates that our model provides a reasonable
match to the O~{\sc i} $\lambda5513$, but is unable to explain the profiles
of O~{\sc i} $\lambda7773$ and N~{\sc i} $\lambda8223$, 
whose widths appear to be broader that those of the predicted lines.
As pointed out by
Sharpee et al. (\cite{sharpee04}), fluorescence excitation may
contribute substantially to the intensity of 
these lines, and the radiation of the lines may originate
predominantly in the outer neutral regions.

\section{Conclusions}

This paper addresses the potential of emission line profiles 
 to probe the physical conditions of PNe.
Our results show that if temperature or density variations are
present within the nebulae,
lines with different excitation temperatures or critical densities
might show significantly different profiles 
even though they originate in similar
 ionic species. We show that line profiles provide a new
way to investigate the CEL/ORL abundance-discrepancy problem. 
For this purpose, we require high-resolution and deep spectroscopic data.
In the ideal case, the line width contributed
by instrumental broadening
should be smaller than that by the thermal broadening of heavy-element lines.
This technical requirement corresponds to
a spectral resolution higher than $75,000$.

We present the electron temperatures and densities in the velocity space of
a sample of PNe. No significant temperature or density variations
are found.
This provides a lower limit to the temperature and density inhomogeneities
along the line of sight.

We attempt to reproduce the observational data of two PNe, NGC~6153 and NGC~7009,
acquired by Barlow et al. (\cite{barlow06}), which
show that the [O~{\sc iii}] $\lambda5007$,
the [O~{\sc iii}] $\lambda4363$, and the O~{\sc ii} recombination lines
have different profiles.
We find that a pure photoionization model of chemically-homogeneous gas can
explain the [O~{\sc iii}] $\lambda5007$/$\lambda4363$ profile discrepancies,
but cannot explain the [O~{\sc iii}]/O~{\sc ii} profile discrepancies.
We thus conclude that CEL and ORL may originate in kinematically-different 
nebular
components. Alternatively, the [O~{\sc iii}]/O~{\sc ii} profile discrepancies
could be caused by extra heating in the outer regions.
In this work, 1D computations were completed. Line profiles
may significantly depend on
nebular geometrical structures, as proposed by
Morisset \& Stasi\'{n}ska (\cite{morisset06a}).
However, for lines from the same ionic species,
nebular geometrical structures provide an effect on the profiles
of a similar degree, and thus hardly affect our results.

We construct line profiles of the approximatively-spherical PN, IC~418.
Our model can explain most of the observed line profiles.
A velocity field that sharply increases
outwards is revealed.
We find that the [S~{\sc ii}] and [O~{\sc ii}]
density diagnostic lines have very different profiles, which imply
that large density variations are present within the PN. Our model
shows that in the low-velocity regions of the
[O~{\sc i}] $\lambda6300$ line, the predicted flux is lower than
the observed one, which indicates the existence of neutral clumps within
the ionized regions. Generally, the profiles of CELs and ORLs are in good
agreement. However, IC~418 is a young PN which has a small CEL/ORL
abundance-discrepancy, and thus is unsuitable for studying the
abundance problem.
For further studies, high-quality spectroscopic observations
of PNe with large CEL/ORL abundance-discrepancies should be invaluable.

\begin{acknowledgements}

I am grateful to Robert Willimas for obtaining
the echelle spectra of IC~418 from their Sharpee et al. (\cite{sharpee04})
paper. I am indebted to Xiao-Wei Liu and Sun Kwok for 
valuable discussions.
I also thank the referee Grazyna Stasi\'{n}ska for
essential suggestions that helped to clarify and improve
this paper significantly. Financial support from HKU is
acknowledged.

\end{acknowledgements}

\end{document}